\documentclass[12pt]{article}

\usepackage[a4paper,left=30mm,right=20mm,top=20mm,bottom=20mm]{geometry}
\usepackage{graphics}
\usepackage{epsfig}
\usepackage{cite}

\title{TMD parton densities in associated real and virtual photon and jet production at LHC}

\author{A.V.~Lipatov$^{1,\,2}$, N.P.~Zotov$^1$}

\begin{document}

\maketitle

\begin{center}

{\it $^1$Skobeltsyn Institute of Nuclear Physics, Lomonosov Moscow State University, 119991 Moscow, Russia}\\
{\it $^2$Joint Institute for Nuclear Research, 141980 Dubna, Moscow Region, Russia}

\end{center}

\vspace{0.5cm}

\begin{center}

{\bf Abstract }

\end{center}

We study the associated production of real (isolated) or virtual photons 
(with their subsequent leptonic decay) and hadronic jets in proton-proton collisions at the LHC
using the $k_T$-factorization approach of QCD.
The consideration is based on the 
off-shell quark-gluon QCD Compton scattering 
subprocesses. In the case of virtual photon production, the contributions from $Z$ boson 
exchange as well as $\gamma^* - Z$ interference with the full spin correlations are included. 
The transverse momentum dependent (TMD) quark and gluon densities in a proton are determined from the 
Kimber-Martin-Ryskin prescription or Catani-Ciafoloni-Fiorani-Marchesini (CCFM) equation. 
In the latter, we restricted to the case where the gluon-to-quark splitting occurs at the last evolution 
step and calculate the sea quark density as a convolution of the CCFM-evolved gluon distribution and the
TMD gluon-to-quark splitting function.
Our numerical predictions are compared with the recent experimental data taken by the ATLAS
Collaboration. We discuss the theoretical uncertainties of our calculations
and argue that further studies are capable of constraining the
TMD parton densities in a proton.

\vspace{1.0cm}

\noindent
PACS number(s): 12.38.-t, 12.15.Ji

\newpage

Recently, the ATLAS Collaboration has reported data\cite{1,2} on the
associated direct photon\footnote{Usually the photons are called direct if they are coupled to the
interacting quarks.} and hadronic jet production
in proton-proton collisions at the LHC.
The data on the associated production of lepton pair and jets have been presented also\cite{3}.
At present, both these processes are subjects of intense studies.
The theoretical and experimental investigations of
direct photon production provide a probe of the hard subprocess dynamics since the produced
photons are almost insensitive to the effects of final-state hadronization.
The corresponding total and differential cross sections are strongly sensitive to the parton (quark and gluon)
content of a proton since, at leading order, the direct photons can be produced mainly via quark-gluon
Compton scattering or quark-antiquark annihilation.
Moreover, such events provide one of
the main backgrounds in searches of Higgs bosons decaying to a photon pair.
Dilepton production, where final leptons originate from the decay of virtual photon 
or intermediate $Z$ boson, 
has a large cross section and clean
signature in the detectors and therefore it is used for monitoring the collider luminosity and
calibration of detectors. It is an important reference process for measurements of
electroweak boson properties and provides a major source of 
background to a number of processes such as Higgs, $t\bar t$ pair, di-boson or $W^\prime$ 
and $Z^\prime$ bosons production (and other processes beyond the Standard Model)
studied at hadron colliders.

It was claimed\cite{3} that recent ATLAS data on the associated dilepton and hadronic jet 
production can be reasonably well described by the next-to-leading perturbative QCD predictions (NLO pQCD) 
computed using the \textsc{BlackHat} program\cite{4}. 
The NLO pQCD calculations\cite{5} (the \textsc{JetPhox} package) provide also a 
reasonably good description of the ATLAS data\cite{1,2}
on the associated direct photon and jet production, except for the case of
azimuthal opening angle between the produced photon and jet.
Additionally, it was demonstrated\cite{1} that the theoretical predictions\cite{5} overestimate 
the measured cross sections at small photon transverse energy $E_T^\gamma < 45$~GeV.

An alternative description can be achieved within the framework of the $k_T$-factorization QCD approach\cite{6,7}. 
This approach is based on the Balitsky-Fadin-Kuraev-Lipatov (BFKL)\cite{8}
or the Ciafaloni-Catani-Fiorani-Marchesini (CCFM)\cite{9} gluon dynamics
and provides solid theoretical grounds for the effects of initial gluon radiation
and intrinsic parton transverse momentum\footnote{A detailed description and discussion of the
$k_T$-factorization formalism can be found in\cite{10}.}.
The basic dynamical quantities of the $k_T$-factorization formalism are the parton distributions 
unintegrated over the parton transverse momentum $k_T$, or 
transverse momentum dependent (TMD) parton distributions. 
At present, these quantities are subject of intense studies, and
various theoretical approaches to investigate them
have been proposed\cite{11,12,13,14,15}. 
Nevertheless, most of phenomenological applications 
take only gluon and valence quark contributions 
into account (see, for example,\cite{16,17,18,19,20}). Such approach is a reasonable 
approximation (based on the 
dominance of spin-$1$ exchange processes at high energies) for the production 
processes coupled to the gluons. But, to correctly treat the final states 
associated with the quark-initiated processes, it is necessary to go beyond 
this approximation and take into account subleading effects connected, in particular, 
with the TMD sea quark distribution. First attempts to address this issue have been performed in\cite{21,22,23,24}, 
where the TMD sea quark density has been derived from the TMD gluon distribution 
via splitting probabilities to lowest order of perturbative theory, neglecting any transverse
momentum dependence in the gluon-to-quark branching.
Recently, the formulation of the TMD sea quark density
which incorporates the effects of the TMD gluon-to-quark splitting 
function\cite{25} have been proposed\cite{26}, where 
the TMD gluon-to-quark splitting 
function contains all single logarithmic small-$x$
corrections to sea quark evolution for any order of perturbation
theory. The proposed formulation has been implemented in
a Monte Carlo event generator \textsc{cascade}\cite{27}, and the
specific kinematical effects from initial state parton transverse momentum on 
the forward $Z$ boson spectrum have been studied\cite{28}.
First phenomenological application of the developed formalism\cite{26,28}
to the analysis of experimental data was made in\cite{29},
where the inlcusive Drell-Yan lepton pair production at the LHC
has been considered. In the present note we extend this previous investigation 
by including into the consideration the processes of 
associated production of direct photons or lepton pairs and hadronic jets at the LHC. 
As it was mentioned above, both these processes offer high sensitivity to the sea quark 
evolution in a proton at moderate and high scales (up to $\mu^2 \sim m_Z^2$). 

Let us start from a short review of calculation steps.
Our consideration is based on the
off-shell quark-gluon QCD Compton-like scattering 
subprocesses\footnote{We will neglect the contributions from the so-called 
fragmentation mechanisms in the case of direct photon production. It is because after
applying the isolation cut (see\cite{1,2}) these contributions amount only
to about 10\% of the visible cross section.
The isolation requirement and additional conditions which preserve our 
calculations from divergences have
been specially discussed in\cite{30}.}:
$$
  q(k_1) + g^*(k_2) \to \gamma(p_1) + q(p_2),\eqno(1)
$$
$$
  q(k_1) + g^*(k_2) \to Z/\gamma^* + q \to l^+(p_1) + l^-(p_2) + q(p_3),\eqno(2)
$$

\noindent
where the four-momenta of all corresponding particles are given in
the parentheses. 
Since we are interested in the events containing the jets in final state,
using the subprocess~(2) istead of simple 
quark-antiquark annihilation
(which has been applied previously\cite{29} to the inclusive Drell-Yan production case) 
is more suitable (see discussion below).
Also, we will neglect the virtualities of initial quarks (but not gluons)
in production amplitudes of subprocesses~(1) and~(2) as compared to the quite large hard scale $\mu^2$ of such events.
Note that contributions from the quark-antiquark annihilation
are effectively taken into account in our consideration due to 
initial state gluon radiation. It is in contrast with collinear QCD
factorization where these contributions have to be taken into account 
separately.

The gauge-invariant off-shell production amplitudes squared of subprocesses~(1) and~(2)
have been calculated in\cite{31} and\cite{32}, respectively. 
These calculations are rather straightforward. We only mention that, 
in according to the $k_T$-factorization prescription\cite{6,7}, 
the summation over the incoming off-shell gluon polarizations is carried out with
$\sum \epsilon^\mu \epsilon^{*\,\nu} = {\mathbf k}_{2T}^\mu {\mathbf k}_{2T}^\nu / {\mathbf k}_{2T}^2$,
where ${\mathbf k}_{2T}$ is the gluon transverse momentum, and
$k_2^2 = - {\mathbf k}_{2T}^2 \neq 0$. In all other respects our
calculations follow the standard Feynman rules.

According to the $k_T$-factorization approach, to calculate the cross section of 
processes under consideration one should convolute corresponding off-shell partonic 
cross sections with the TMD parton densities in a proton.
Our master formulas read:
$$
  \displaystyle \sigma(pp \to \gamma + {\rm jet})=\sum_q \int\frac{1}{16\pi\, (x_1 x_2 s)^2} \,|\bar {\mathcal M}(qg^* \to \gamma q)|^2 \times \atop { 
    \displaystyle \times f_q(x_1,\mathbf k_{1T}^2,\mu^2) f_{g}(x_2,\mathbf k_{2T}^2,\mu^2) d\mathbf p_{1T}^2 d\mathbf k_{1T}^2 d\mathbf k_{2T}^2 dy_1 dy_2 \frac{d\phi_1}{2\pi}\frac{d\phi_2}{2\pi} }, \eqno(3)
$$
$$
  \displaystyle \sigma(pp \to l^+ l^- + {\rm jet})=\sum_q \int\frac{1}{256\pi^3\, (x_1 x_2 s)^2} \,|\bar {\mathcal M}(qg^* \to Z/\gamma^* q \to l^+ l^- q)|^2 \times \atop { 
    \displaystyle \times f_q(x_1,\mathbf k_{1T}^2,\mu^2) f_{g}(x_2,\mathbf k_{2T}^2,\mu^2) d\mathbf p_{1T}^2 d\mathbf p_{2T}^2 d\mathbf k_{1T}^2 d\mathbf k_{2T}^2 dy_1 dy_2 dy_3\frac{d\phi_1}{2\pi}\frac{d\phi_2}{2\pi} \frac{d\psi_1}{2\pi} \frac{d\psi_2}{2\pi}}, \eqno(4)
$$
\noindent
where $f_q(x,\mathbf k_{T}^2,\mu^2)$ and $f_{g}(x,\mathbf k_{T}^2,\mu^2)$ are the TMD quark and gluon densities
in a proton, $s$ is the total energy, $\mathbf p_{1T}$, $\mathbf p_{2T}$, $\psi_1$, $\psi_2$, $y_1$, $y_2$
and $y_3$ are the transverse momenta, azimuthal angles and center-of-mass rapidities
of produced particles, and $\phi_1$ and $\phi_2$ are the azimuthal angles of the incoming partons
having the non-zero transverse momenta $\mathbf k_{1T}$ and $\mathbf k_{2T}$ and
fractions $x_1$ and $x_2$ of the longitudinal momenta of the colliding protons.
If we average these expressions over $\phi_1$ and $\phi_2$ and take the limit
$|\mathbf k_{1T}| \to 0$ and $|\mathbf k_{2T}| \to 0$, then we recover the corresponding formulas 
of the collinear QCD factorization.

To calculate the TMD parton densities in a proton we follow the approach\cite{26} based on the 
CCFM equation. 
The CCFM parton
shower describes only the emission of gluons,
while real quark emissions are left aside. 
It implies that this 
equation describes only the distinct evolution of TMD gluon and valence quarks, 
while the non-diagonal transitions between quarks and gluons are absent.
The TMD gluon and valence quark distributions have been obtained from the 
numerical solutions of the CCFM equation in\cite{17,33}.
To calculate the TMD sea quark density we apply the approximation where the sea quarks 
occur in the last gluon-to-quark splitting. 
At the next-to-leading logarithmic
accuracy $\alpha_s (\alpha_s \ln x)^n$ the TMD sea quark distribution can be written as follows\cite{26}:
$$
  f_q^{\rm (sea)}(x,{\mathbf q}_T^2,\mu^2) = \int \limits_x^1 {dz \over z} \int d{\mathbf k}_T^2
    {1\over {\mathbf \Delta}^2} {\alpha_s \over 2\pi} P_{qg}(z,{\mathbf k}_T^2,{\mathbf \Delta}^2) f_g(x/z,{\mathbf k}_T^2, \bar \mu^2),\eqno(5)
$$

\noindent
where $z$ is the fraction of the gluon light cone momentum carried out by
the quark, ${\mathbf q}_T$ and $\mathbf k_T$ are the sea
quark and gluon transverse momenta, $z$ is the fraction of the gluon light cone momentum carried out by
the quark, and $\mathbf \Delta = {\mathbf q}_T - z{\mathbf k}_T$. 
The sea quark evolution is driven\footnote{In\cite{30} the TMD sea quark contributions have 
been simulated using the off-shell gluon-gluon fusion subprocess.}
by the off-shell gluon-to-quark
splitting function $P_{qg}(z,{\mathbf k}_T^2,{\mathbf \Delta}^2)$\cite{25}:
$$
  P_{qg}(z,{\mathbf k}_T^2,{\mathbf \Delta}^2) = T_R \left({\mathbf \Delta}^2\over {\mathbf \Delta}^2 + z(1-z)\,{\mathbf k}_T^2\right)^2
    \left[(1 - z)^2 + z^2 + 4z^2(1 - z)^2 {{\mathbf k}_T^2\over {\mathbf \Delta}^2} \right],\eqno(6)
$$

\noindent 
with $T_R = 1/2$. The splitting function $P_{qg}(z,{\mathbf k}_T^2,{\mathbf \Delta}^2)$
has been obtained by generalizing to finite transverse momenta, in the high-energy region, 
the two-particle irreducible kernel expansion\cite{34}. Although evaluated off-shell, 
this splitting function is universal: it
takes into account the small-$x$ enhanced transverse momentum dependence 
up to all orders in the strong coupling, and reduces to the conventional splitting
function at lowest order for $|\mathbf k_{T}| \to 0$.
The scale $\bar \mu^2$ is defined\cite{26} from the angular ordering condition which is natural
from the point of view of the CCFM evolution: $\bar \mu^2 = {\mathbf \Delta}^2/(1-z)^2 + {\mathbf k}_T^2/(1-z)$.
To precise, in~(5) we have used A0 gluon\cite{33}.

An alternative way to calculate the TMD parton densities
in a proton is the Kimber-Martin-Ryskin (KMR) approach\cite{35},
which is a formalism to construct the TMD 
parton distributions from the known collinear ones.
The key assumption of the KMR approach is that the
$k_T$-dependence of the TMD parton densities
enters at the last evolution step, and the conventional 
Dokshitzer-Gribov-Lipatov-Altarelli-Parisi (DGLAP) evolution equations\cite{36} can 
be used up to this step.
The TMD parton densities calculated using both these approaches have been 
compared in\cite{29}. Below we will test them numerically\footnote{We have used the leading-order 
MSTW'2008 parton densities\cite{37} as input in the KMR prescription.}.

Other essential parameters were taken as follows: renormalization 
and factorization scales $\mu_R = \mu_F = \xi E_T^\gamma$ or
$\mu_R = \mu_F = \xi M$, where $E_T^\gamma$ and $M$ are the final photon 
transverse energy and invariant mass of produced lepton pair, respectively.
We vary the parameter $\xi$ between $1/2$ and $2$ about the default value
$\xi = 1$ in order to estimate the scale uncertainties of our calculations.
Next, following to\cite{38}, we set $m_Z = 91.1876$~GeV, $\Gamma_Z = 2.4952$~GeV,
$\sin^2 \theta_W = 0.23122$ and use the LO formula for the strong coupling constant
$\alpha_s(\mu^2)$ with $n_f= 4$ active quark flavors at
$\Lambda_{\rm QCD} = 200$~MeV, so that $\alpha_s (m_Z^2) = 0.1232$.
Since we investigate a wide region of $E_T^\gamma$ and $M$, we use the running QED coupling
constant $\alpha(\mu^2)$. To take into account the non-logarithmic loop corrections to the
dilepton production cross section we apply the effective $K$-factor,
as it was done in\cite{39}:
$$
  K = \exp \left[ C_F {\alpha_s(\mu^2)\over 2\pi} \pi^2 \right],\eqno(7)
$$

\noindent
where color factor $C_F = 4/3$. A particular scale choice
$\mu^2 = {\mathbf p}_T^{4/3} M^{2/3}$
(with ${\mathbf p}_T$ being the transverse momentum of
produced lepton pair) has been proposed\cite{39}
to eliminate sub-leading logarithmic terms. Note we choose this scale
to evaluate the strong coupling constant in~(7) only.
Everywhere the multidimensional integration have been performed
by the means of Monte Carlo technique, using the routine \textsc{vegas}\cite{40}.
The corresponding C++ code is available from the authors on request\footnote{lipatov@theory.sinp.msu.ru}.

We now are in a position to present our numerical results.
The ATLAS Collaboration has measured\cite{2} the direct photon plus jet production 
cross sections as a function of the photon
transverse energy $E_T^{\gamma}$, leading jet transverse momentum $p_T^{\rm jet}$ and rapidity $y^{\rm jet}$,
photon-jet invariant mass $M^{\gamma - {\rm jet}}$, difference $\Delta \phi^{\gamma - {\rm jet}}$ 
between the azimuthal angles of the photon and jet
and scattering angle $\cos\theta^*$ in the photon-jet centre-of-mass frame.
In addition, the differential cross section $d\sigma/dE_T^\gamma$ has been measured\cite{1}
for three different rapidity ranges of leading jet:
$|y^{\rm jet}| < 1.2$, $1.2 < |y^{\rm jet}| < 2.8$ and $2.8 < |y^{\rm jet}| < 4.4$.
For each rapidity configuration the same-sign ($\eta^\gamma y^{\rm jet} > 0$)
and opposite-sign ($\eta^\gamma y^{\rm jet} < 0$) cases have been studied separately,
where $\eta^\gamma$ is the produced photon pseudo-rapidity.
The differential cross sections of associated
$Z/\gamma^* \to l^+ l^-$ 
and jet production were
measured\cite{3} as a function of the jet transverse momentum $p_T^{\rm jet}$ and rapidity $y^{\rm jet}$
at $66 < M < 116$~GeV, $p_T^l > 20$~GeV, $|\eta^l| < 2.5, $ $p_T^{\rm jet} > 20$~GeV and $|y^{\rm jet}| < 4.4$.
These measurements were also performed as a function of the dijet invariant mass $M^{\rm jet-jet}$
and angular separation $\Delta \phi^{\rm jet-jet}$ between the two leading jets in events with at 
least two jets in the final state.

To calculate the production rates of both semi-inclusive processes under consideration we
apply the procedure which has been used previously in\cite{41,42,43}. 
The produced photon or lepton pair is accompanied by a number
of partons radiated due to the non-collinear parton evolution. 
From these several jets we choose the one 
carrying the largest transverse energy, and then compute the semi-inclusive
production cross sections.
The results of our calculations are shown in Figs.~1 --- 4 in comparison 
with the ATLAS data\cite{1,2,3}. We discuss first the distributions on the leading jet rapidity. 
One can see that our predictions based on the KMR parton densities disagree with the data 
and tend to underestimate them in the central rapidity region and overestimate the data in the forward one
for both processes under consideration.
The observed disagreement is due to our approximation for the rapidity of partons coming from the
evolution ladder which form a part of final state jets.
It indicates that the full hadron-level Monte-Carlo event 
generator (like as, for example, \textsc{cascade}) is 
needed to investigate these observables\footnote{Very recently, the associated $W^\pm$ + n jets production has been studied\cite{19} 
with \textsc{cascade}.}. 
Such evaluations are out of present short note.
In Fig.~1 we see that the CCFM-based predictions agree with the ATLAS data 
on the $y^{\rm jet}$ distributions within the theoretical uncertainties.
The distributions on the produced photon transverse energy or jet 
transverse momentum agree reasonably well with the ATLAS data, as it is shown in 
Figs.~2 --- 4.
The exceptions are the KMR predictions for the photon transverse energy 
distribution in the forward $y^{\rm jet}$ region and 
both predictions
for the azimuthal angle separation between the photon and jet
(see Figs.~2 and 3), that are
also connected with the approximation applied. However, the 
CCFM-based predictions agree reasonably well with the ATLAS data
on the photon $p_T^\gamma$ spectrum
in a whole rapidity region.
The sensitivity of predicted cross sections to the TMD quark densities
is clearly visible in the $\Delta \phi^{\gamma - {\rm jet}}$ and 
$\Delta \phi^{\rm jet-jet}$ distributions, as it is shown in Figs.~2 and~4.
It coincides with observations of many papers (see, for example,\cite{19,30} and 
references therein).
None of the TMD quark densities under consideration describe well
the $\Delta \phi^{\gamma - {\rm jet}}$ one at $\Delta \phi^{\gamma - {\rm jet}} \sim 0$,
although the CCFM-based predictions agree well with the data on $\Delta \phi^{\rm jet-jet}$
distribution.
Our calculations show that the influence of sea quarks 
at low $\Delta \phi^{\gamma - {\rm jet}}$ is significant,
and therefore these measurements can be used to better constrain 
the TMD sea quark distributions. 

To conclude, in the present note we have applied the 
TMD quark and gluon densities calculated using the formalism\cite{25} 
to investigate the associated production of real or virtual photons 
and hadronic jets at the LHC.
This study is an extension of previous one\cite{29}
where the inclusive Drell-Yan lepton pair production
has been investigated.
The formalism\cite{25} is based on the TMD gluon-to-quark splitting function\cite{26}
which contains all single logarithmic small-$x$
corrections to sea quark evolution for any order of perturbation
theory. Despite our approximation in description of jets, we obtained a 
reasonably well agreement between our predictions for the
distributions of final particles on the transverse momenta and recent data\cite{1,2,3}
taken by the ATLAS Collaboration at the LHC.
We demonstrated that studies of such processes provide an important 
information about the TMD parton densities in a proton at moderate
and high scales, up to $\mu^2 \sim m_Z^2$. 
In particular, the sensitivity of predicted cross sections to the TMD quark distributions
is clearly visible in the azimuthal angle correlations
between the produced photons or Drell-Yan lepton pairs and/or jets.
It is important for further investigations of small-$x$ physics at hadron colliders,
in particular, in the direction which concerns the non-linear effects originating
from high parton densities at small $x$.
However, for more detailed analysis of considered semi-inclusive processes,
the full hadron-level Monte-Carlo event generator 
should be used.

{\sl Acknowledgements.}
The authors are grateful to H.~Jung and S.P.~Baranov for very useful discussions and comments.
This research was supported by the FASI of Russian Federation
(grant NS-3042.2014.2), RFBR grant 13-02-01060 and the grant of the 
Ministry of education and sciences
of Russia (agreement 8412).
We are also grateful to DESY Directorate for the
support in the framework of Moscow---DESY project on Monte-Carlo implementation for
HERA---LHC.

\newpage

\begin{figure}
\begin{center}
\epsfig{figure=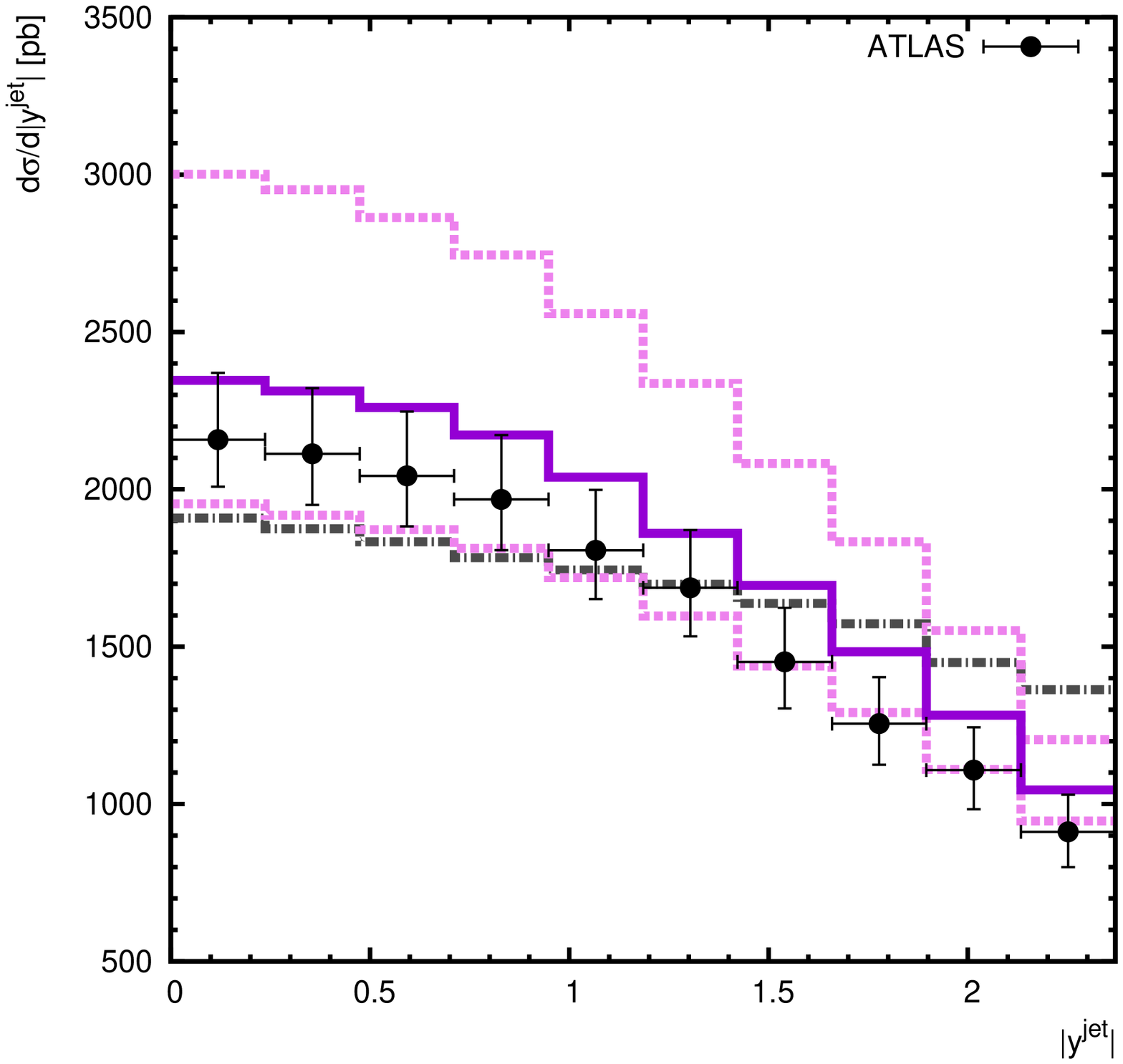, width = 5.5cm, angle = 270}
\epsfig{figure=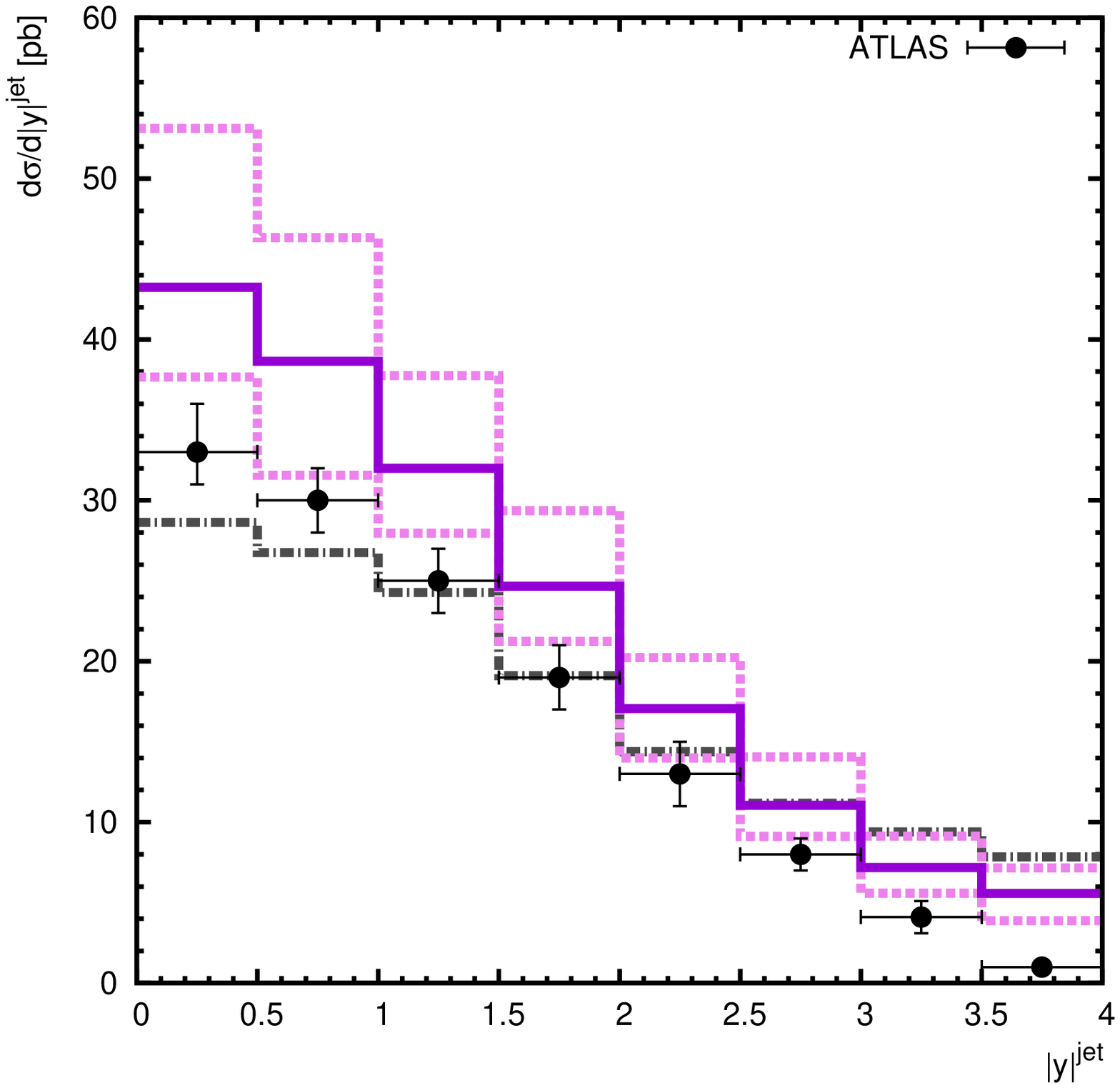, width = 5.5cm, angle = 270}
\end{center}
\caption{The differential cross sections of associated direct photon (left panel) or 
lepton pair (right panel) and jet production in $pp$ collisions at the LHC
as a function of leading jet rapidity. 
The solid and dash-dotted histograms correspond to the CCFM-based and KMR predictions,
respectively. The upper and lower dashed histograms correspond to the scale variations in the
CCFM calculations, as it is described in the text. The experimental data are from ATLAS\cite{2,3}.}
\label{fig1}
\end{figure}

\newpage

\begin{figure}
\begin{center}
\epsfig{figure=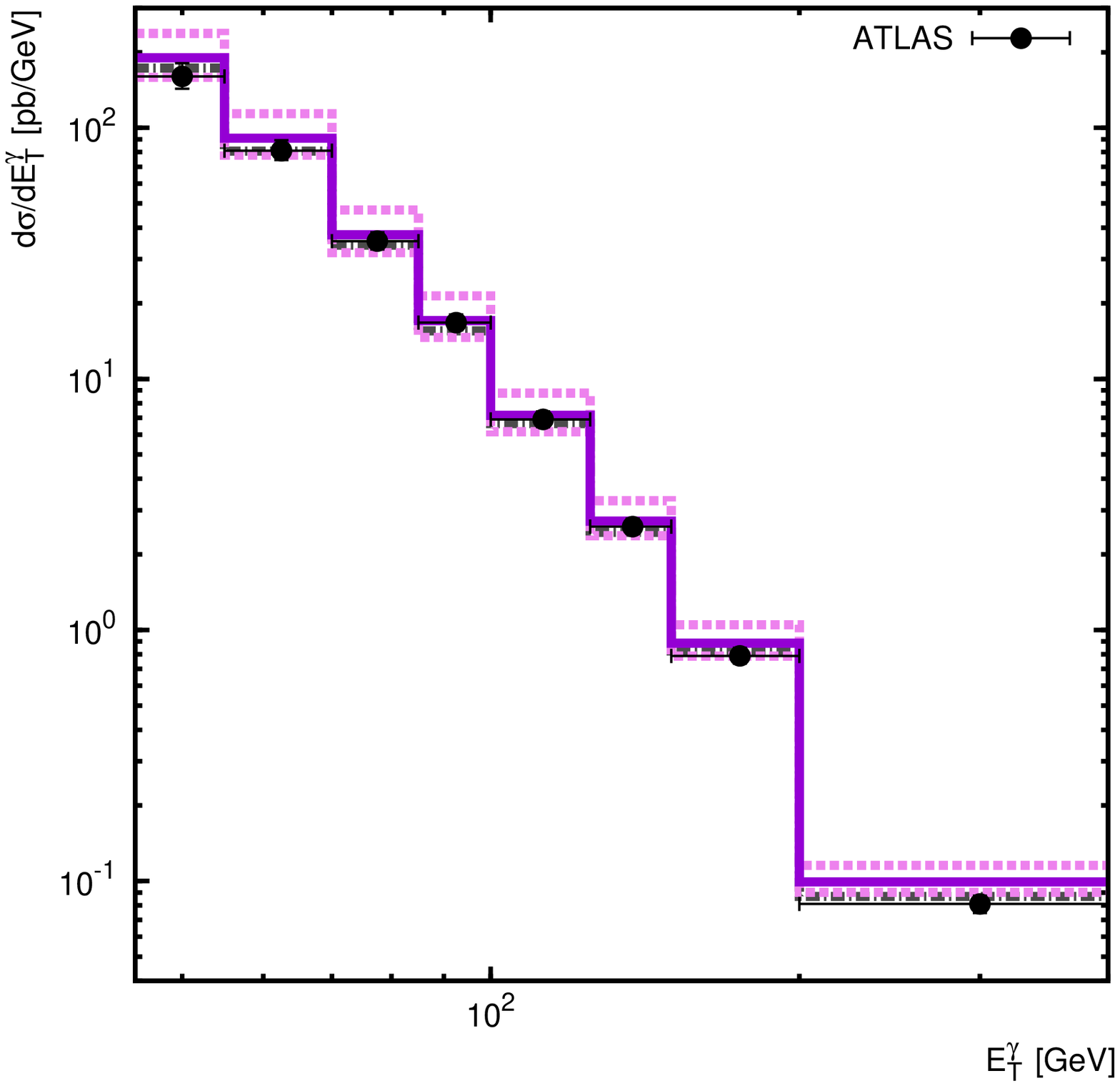, width = 5.5cm, angle = 270}
\epsfig{figure=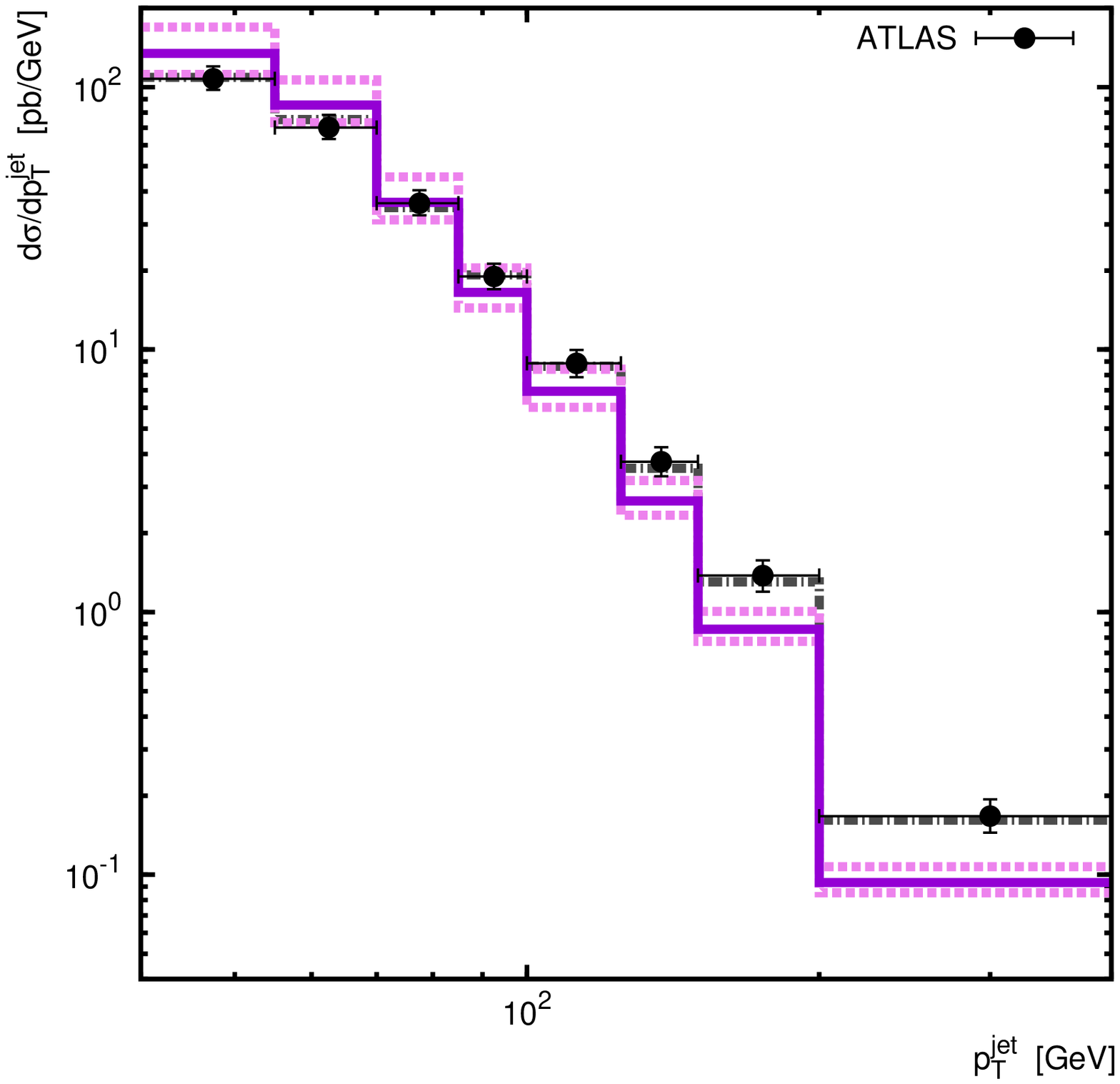, width = 5.5cm, angle = 270}
\epsfig{figure=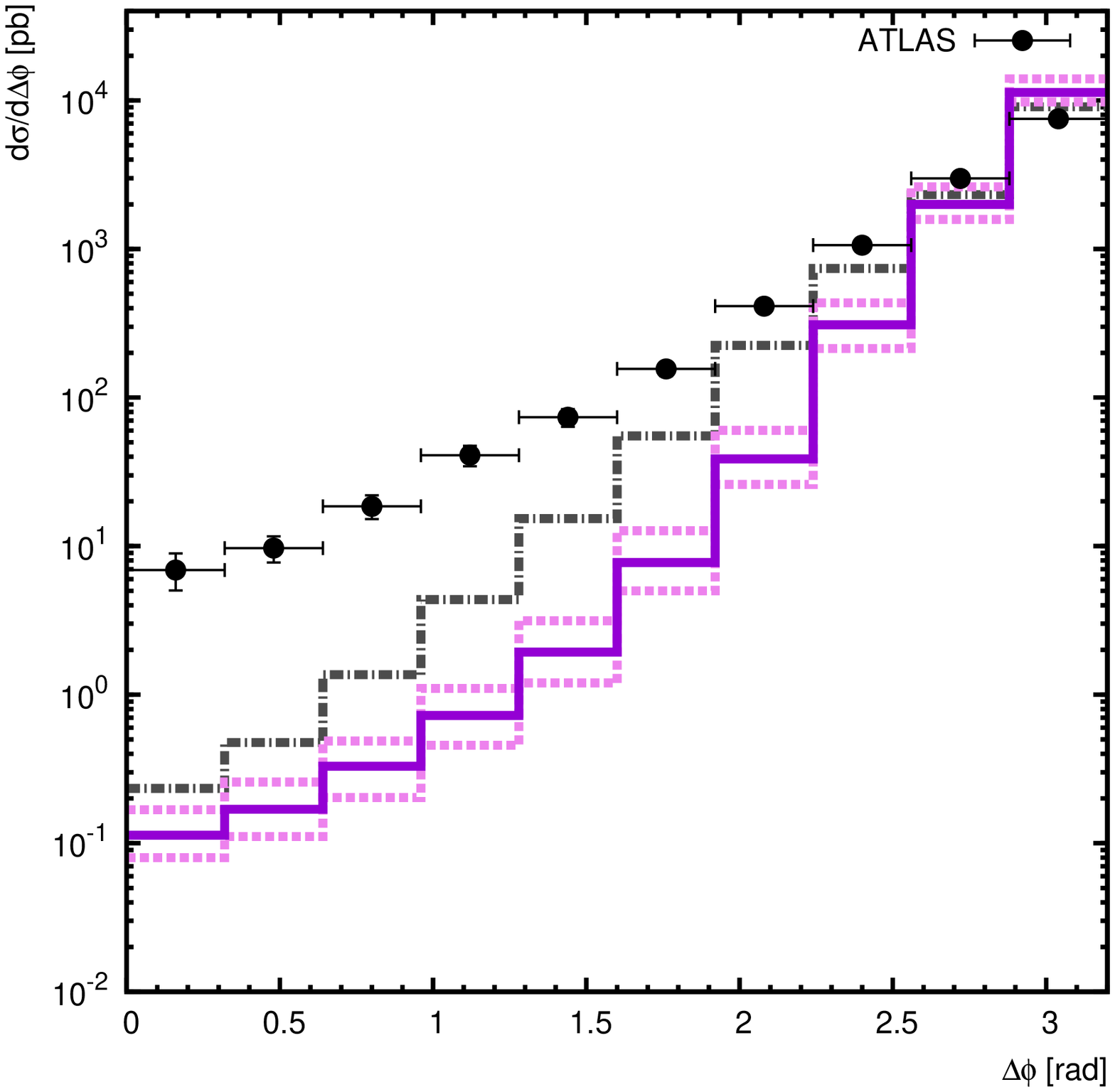, width = 5.5cm, angle = 270}
\end{center}
\caption{The differential cross sections of associated direct photon and jet production in $pp$ collisions
at the LHC. Notation of histograms is the same
as in Fig.~1. The experimental data are from ATLAS\cite{2}.}
\label{fig2}
\end{figure}

\newpage

\begin{figure}
\begin{center}
\epsfig{figure=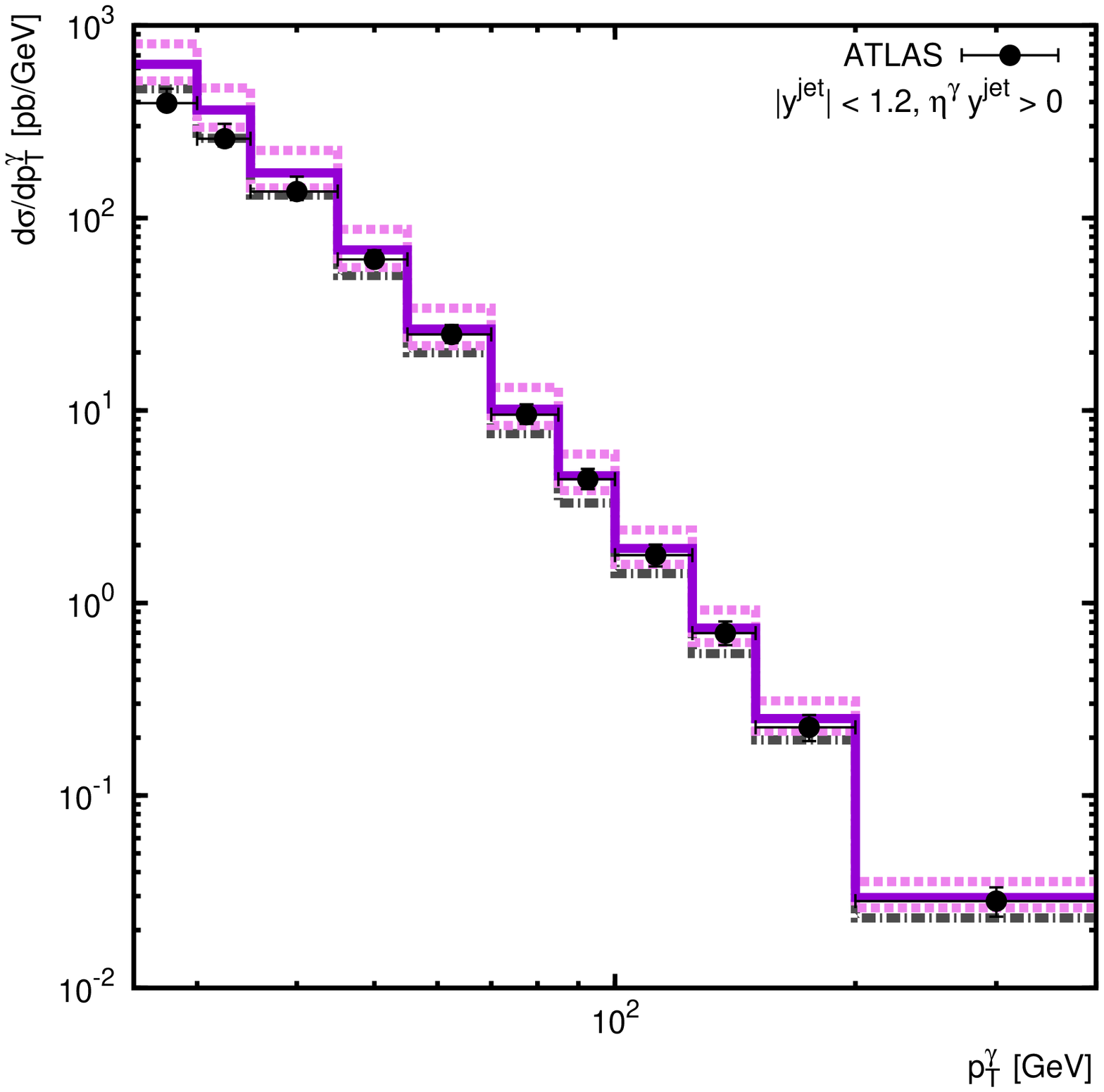, width = 5.5cm, angle = 270}
\epsfig{figure=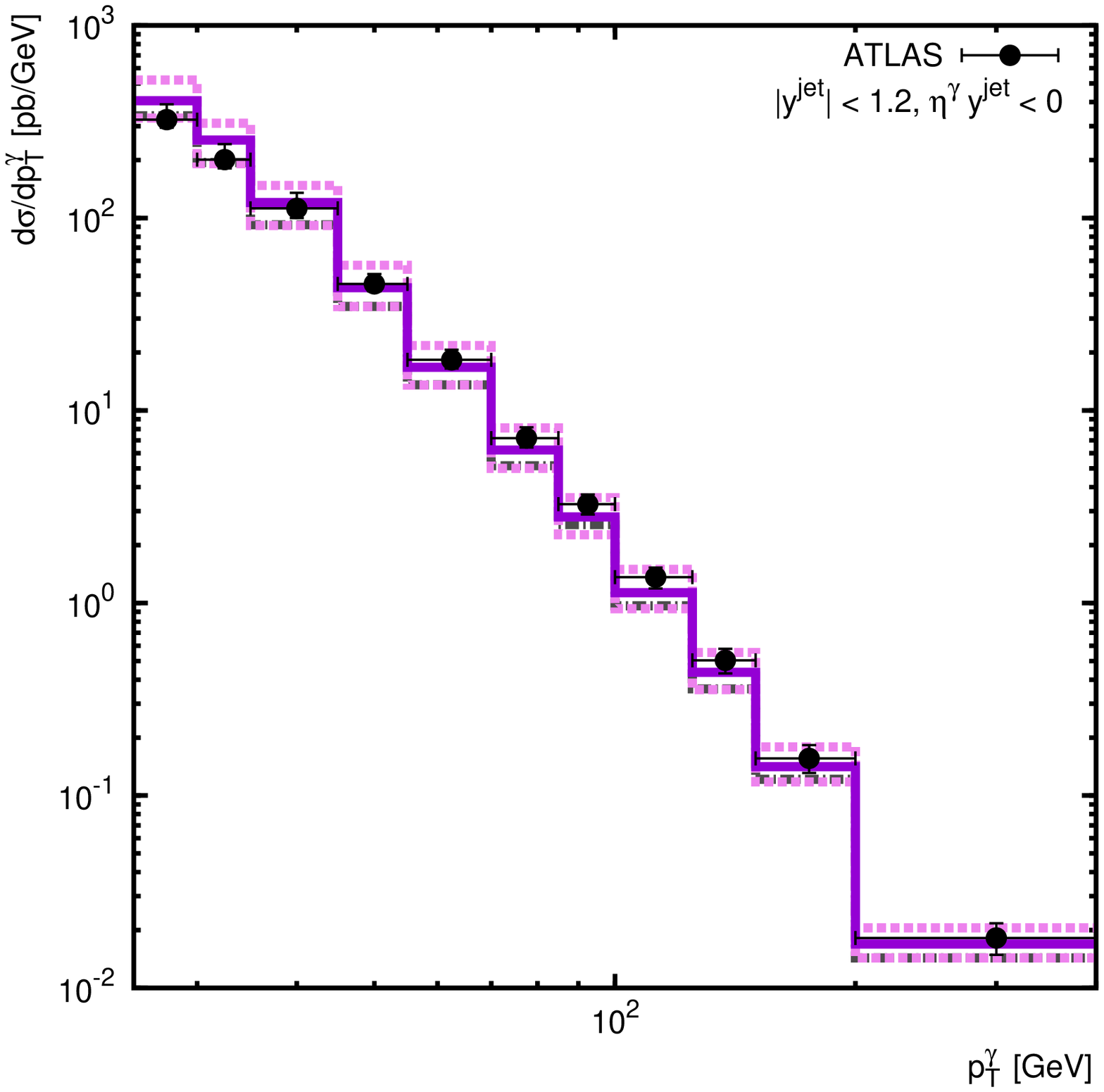, width = 5.5cm, angle = 270}
\epsfig{figure=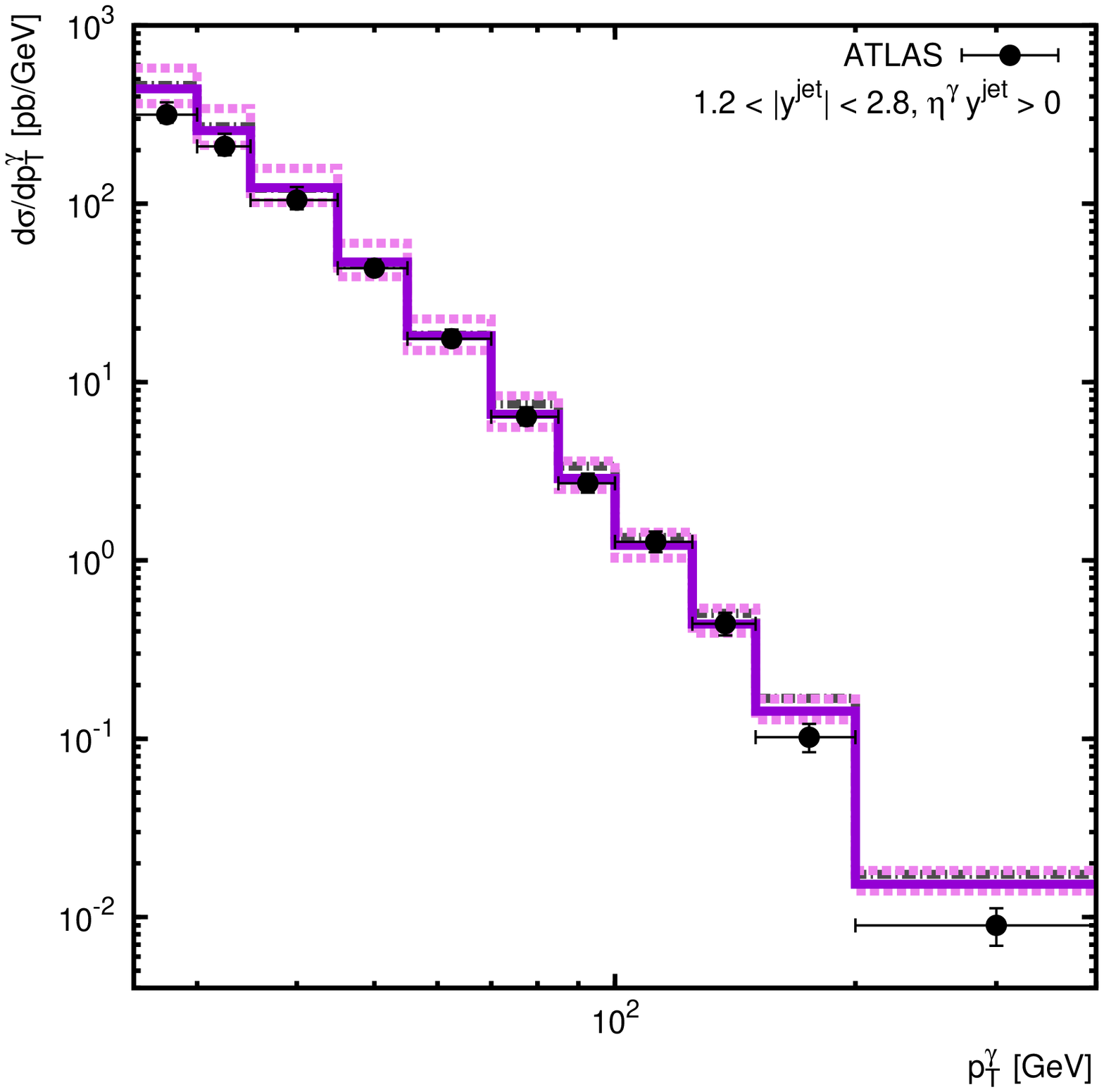, width = 5.5cm, angle = 270}
\epsfig{figure=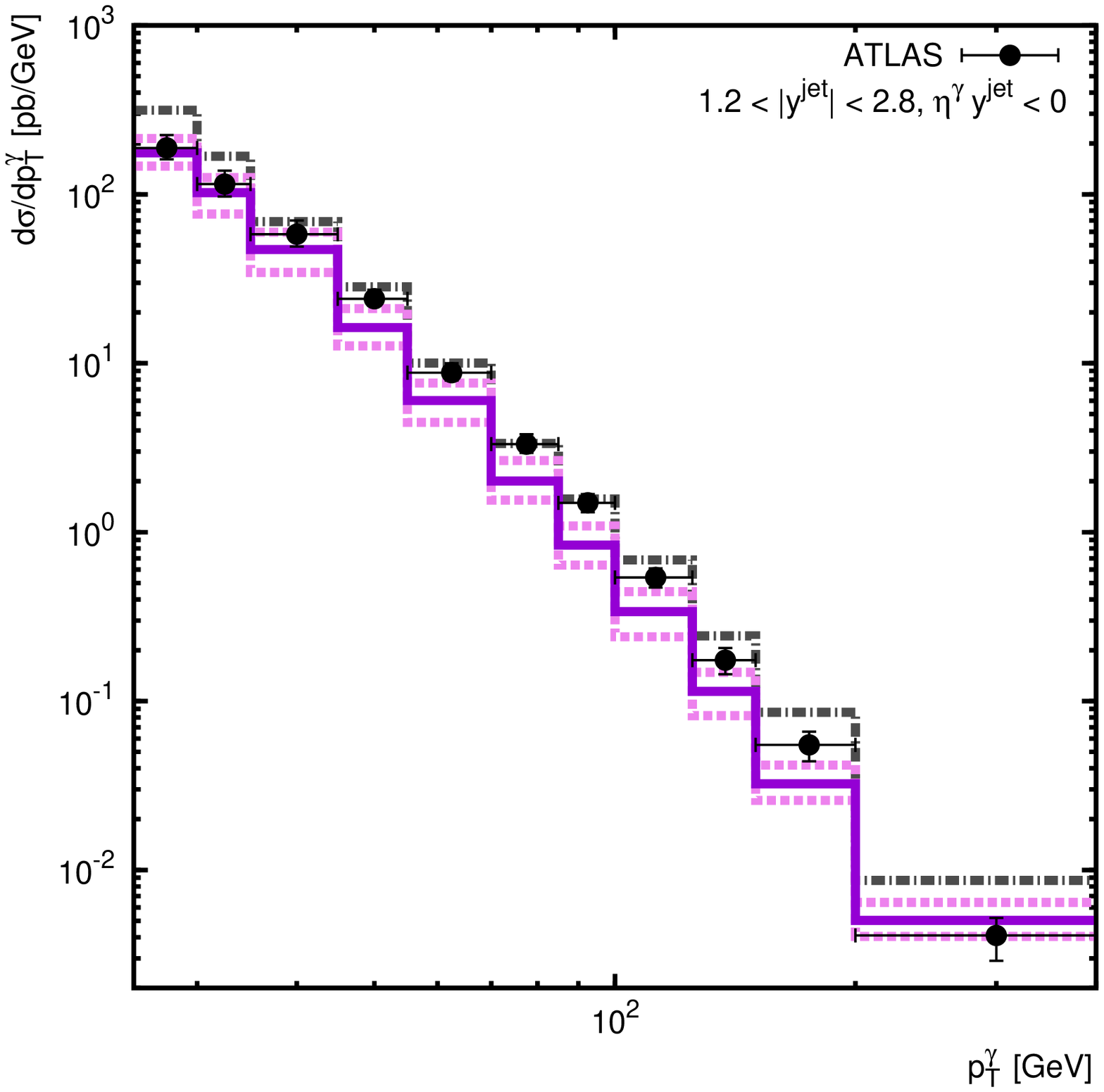, width = 5.5cm, angle = 270}
\epsfig{figure=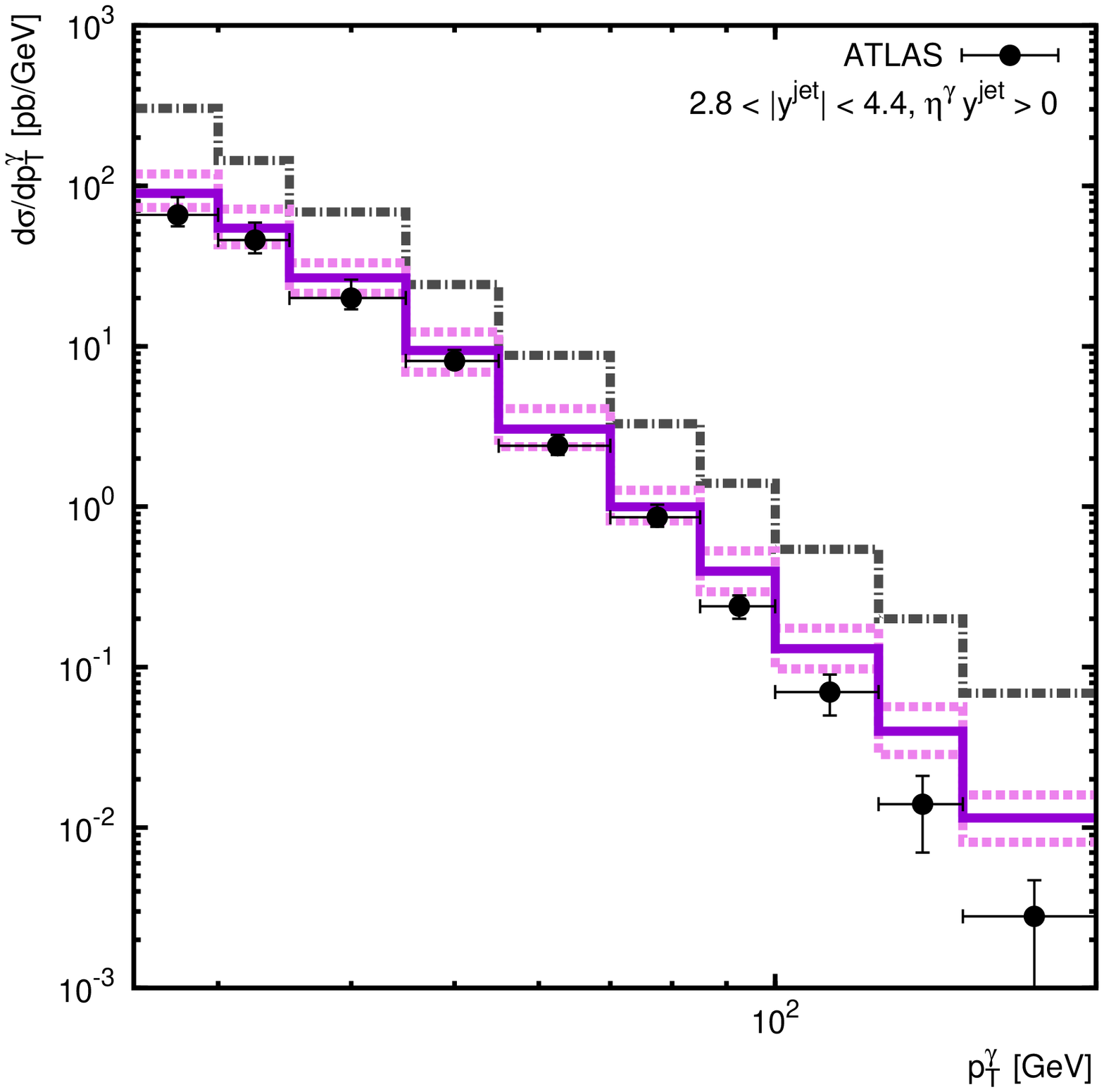, width = 5.5cm, angle = 270}
\epsfig{figure=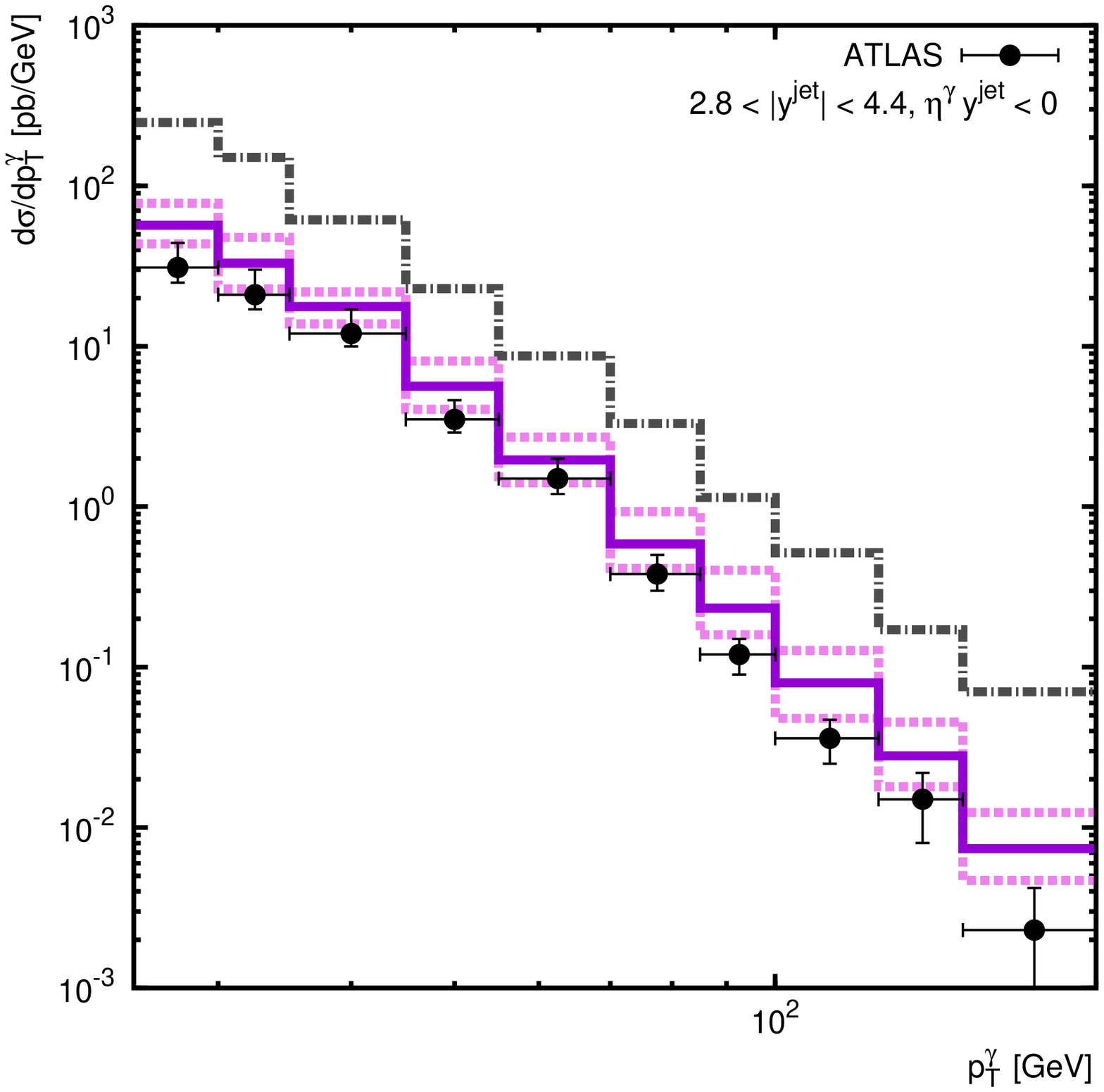, width = 5.5cm, angle = 270}
\end{center}
\caption{The differential cross sections of associated direct photon and jet production in $pp$ collisions
at the LHC. Notation of histograms is the same
as in Fig.~1. The experimental data are from ATLAS\cite{1}.}
\label{fig3}
\end{figure}

\newpage

\begin{figure}
\begin{center}
\epsfig{figure=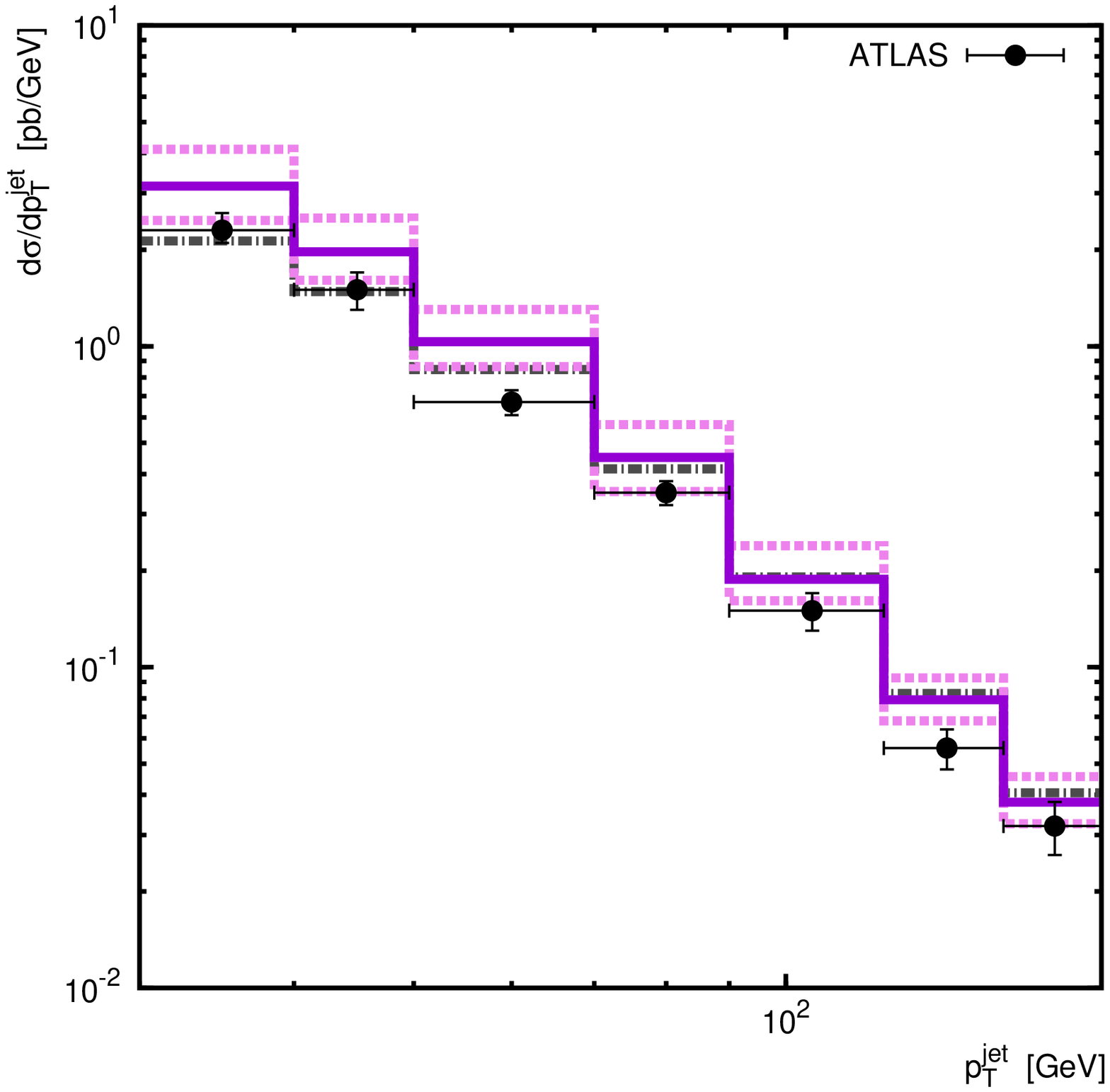, width = 5.5cm, angle = 270}
\epsfig{figure=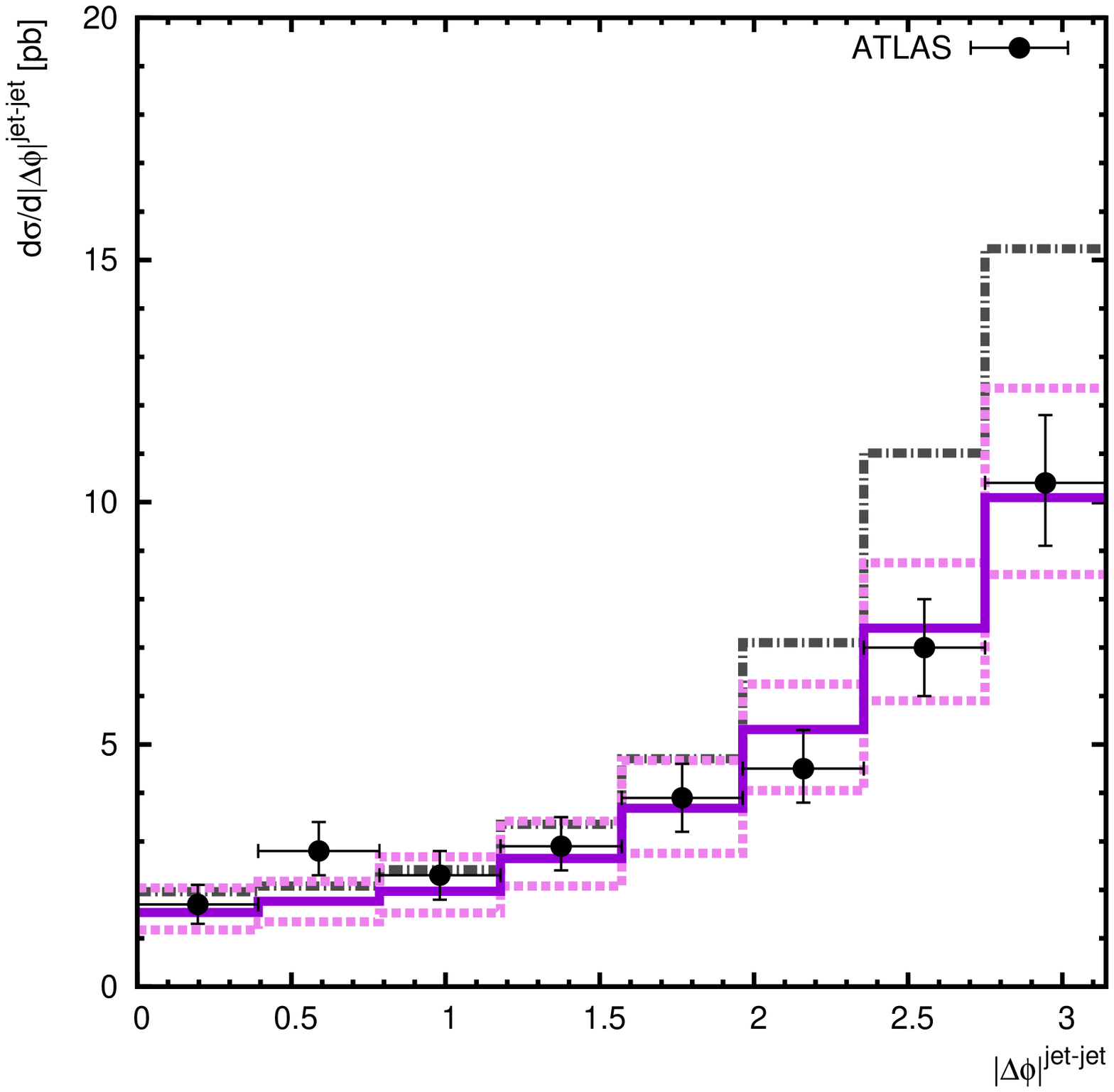, width = 5.5cm, angle = 270}
\end{center}
\caption{The differential cross sections of associated lepton pair and jet production in $pp$ collisions
at the LHC. Notation of histograms is the same
as in Fig.~1. The experimental data are from ATLAS\cite{3}.}
\label{fig4}
\end{figure}

\end{document}